\documentclass[10pt,conference]{IEEEtran}
\usepackage[pdftex]{graphicx}
\usepackage[usenames,dvipsnames,table,xcdraw]{xcolor}
\usepackage{array}
\usepackage{multirow}
\usepackage[normalem]{ulem}
\useunder{\uline}{\ul}{}
\usepackage{booktabs}
\usepackage{supertabular,booktabs}
\usepackage{longtable}
\usepackage{lscape}
\usepackage{cite}
\usepackage{url}
\usepackage{dirtytalk}
\usepackage{graphicx}
\usepackage{comment}
\usepackage{makecell}
\usepackage{svg}
\usepackage{amsmath}
\usepackage{balance}
\usepackage[caption=false]{subfig}
\graphicspath{{Images/}}
\usepackage[normalem]{ulem}
\useunder{\uline}{\ul}{}

\usepackage{tabularx}
\usepackage{enumitem}
\usepackage{calc}
\usepackage{parskip}

\usepackage[most]{tcolorbox}

\newtcolorbox{mytextbox}[1][]{%
  sharp corners,
  enhanced,
  colback=white,
  attach title to upper,
  #1
}

\newcommand{\numPrimaryStudies}{243}
\newcommand{\numQualitative}{36}
\newcommand{\totalRetrieved}{961}
\newcommand{\numTitleAbScreening}{857}
\newcommand{\numTAScreeningExcluded}{513}
\newcommand{\numFullTextScreening}{344}
\newcommand{\numFTScreeningExcluded}{101}
\newcommand{\numDuplicateManual}{46}
\newcommand{\numDuplicateAuto}{58}

\begin{document}

\title{Sustainability is Stratified: Toward a Better Theory of Sustainable Software Engineering}

\author{\IEEEauthorblockN{Sean McGuire, Erin Schultz, Bimpe Ayoola, and Paul Ralph}
\IEEEauthorblockA{\textit{Faculty of Computer Science},
\textit{Dalhousie University},
Halifax, NS, Canada \\
sn475372@dal.ca, er825179@dal.ca, bm295344@dal.ca, paulralph@dal.ca}
}

%{Shell \MakeLowercase{\textit{et al.}}: Bare Demo of IEEEtran.cls for Computer Society Journals}

\IEEEtitleabstractindextext{%
\begin{abstract}
\textit{Background}: Sustainable software engineering (SSE) means creating software in a way that meets present needs without undermining our collective capacity to meet our future needs. It is typically conceptualized as several intersecting dimensions or ``pillars''---environmental, social, economic, technical and individual. However; these pillars are theoretically underdeveloped and require refinement. 
\textit{Objectives}: The objective of this paper is to generate a better theory of SSE. 
\textit{Method}: First, a scoping review was conducted to understand the state of research on SSE and identify existing models thereof. Next, a meta-synthesis of qualitative research on SSE was conducted to critique and improve the existing models identified. 
\textit{Results}: \totalRetrieved{} potentially relevant articles were extracted from five article databases. These articles were de-duplicated and then screened independently by two screeners, leaving \numPrimaryStudies{} articles to examine. Of these, 109 were non-empirical, the most common empirical method was \textit{systematic review}, and no randomized controlled experiments were found. Most papers focus on ecological sustainability (158) and the sustainability of software products (148) rather than processes. A meta-synthesis of \numQualitative{} qualitative studies produced several key propositions, most notably, that sustainability is \textit{stratified} (has different meanings at different levels of abstraction) and multisystemic (emerges from interactions among multiple social, technical, and sociotechnical systems). 
\textit{Conclusion}: The academic literature on SSE is surprisingly non-empirical. More empirical evaluations of specific sustainability interventions are needed. The sustainability of software development products and processes should be conceptualized as multisystemic and stratified, and assessed accordingly. 
\end{abstract}

}

\maketitle

\IEEEdisplaynontitleabstractindextext

\IEEEpeerreviewmaketitle

\section{Introduction}\label{sec:introduction}

\IEEEPARstart{S}{ustainable} development ``meets the needs of the present without compromising the ability of future generations to meet their own needs''~\cite[p. 29]{brundtland1987report}. Sustainable Software Engineering (SSE), then, means creating software in a way that meets present needs without undermining our collective capacity to meet our future needs. While this sounds simple, some argue that ``computing's dominant frame of thinking is conceptually insufficient to address our current challenges''~\cite{becker2023insolvent}. Indeed, IT systems account for approximately 10\% of global electricity consumption and this proportion is expected to grow~\cite{verdecchia2021green}. The Bitcoin network alone consumes more than 100tWh of electricity per year---more power than many entire nations including Argentina, the Netherlands and the Philippines.\footnote{\url{https://ccaf.io/cbeci/index/comparisons} [accessed 2022/AUG/24]} Incremental improvements in efficiency cannot address such unsustainable resource consumption.

Moreover, while most Software Engineering (SE) researchers consider sustainability in terms of resource consumption and waste reduction~\cite{chitchyan2016sustainability}, Becker et al. argue that software sustainability transcends this uni-dimensional view and encompasses five interdependent dimensions: environmental, social, economic, individual and technical~\cite{becker2015sustainability}. Social sustainability concerns impacts on communities and society, while individual sustainability concerns impacts on \textit{individual} human beings~\cite{penzenstadler2013generic}. Economic sustainability concerns the ability to remain financially viable over an extended period~\cite{penzenstadler2013generic}, while technical sustainability refers to a system's durability and the ease of maintaining and adapting it~\cite{becker2015sustainability, penzenstadler2013generic}. Software is sustainable to the extent that it has a minimal or positive impact on each dimension~\cite{Markus2010Green}.

This five-pillar model has become a notable tool for understanding the impacts of SE~\cite{chitchyan2016sustainability}. However, many scholarly articles on SSE have been published since, and reviewing the literature at this time will help us improve the five-pillar model; that is, better understand and explain the intersection between sustainability and SE. We therefore begin with the following research question.

\begin{mytextbox}
\textbf{Research Question:} What is the current state of research on sustainability in software engineering?
\end{mytextbox}

We address this question by conducting a scoping review and a qualitative meta-synthesis, followed by synthesizing a novel theory of software sustainability.

\section{Related Reviews} \label{sec:relatedwork}

Several reviews of sustainable software and SSE have been conducted previously, and much has been learned from these reviews. For example:

\begin{itemize}
    \item SE has less sustainability research than many similar fields~\cite{penzenstadler2012sustainability}.
    \item Little research assesses sustainability or evaluates sustainability-promoting interventions~\cite{penzenstadler2014systematic,mourao2018green,marimuthu2017software,garcia2018interactions}.
    \item Sustainability is rarely addressed holistically~\cite{gustavsson2020blinded}. Most research on sustainable software focuses on energy efficiency~\cite{mourao2018green,garcia2018interactions,wolfram2017sustainability,CALERO2017117,moises2018practices} or technical sustainability~\cite{anwar2017}. Social factors are sometimes considered important, but never addressed systemically~\cite{gustavsson2020blinded}.
    \item Most research focuses on the sustainability of software \textit{products}, rather than \textit{process}~\cite{marimuthu2017software,wolfram2017sustainability}.
    \item Tool support for sustainability concerns is lacking~\cite{marimuthu2017software}.
    \item Sustainability is complicated and difficult to measure~\cite{calero2013systematic}; metrics for ecological sustainability in software have been neglected~\cite{debbarma2016green}.
    \item Sustainability research tends to suffer from over-complexity, domain-specificity, and lack of general guidance~\cite{penzenstadler2012sustainability}.
    \item Sustainability in \textit{distributed} software development depends on many factors including green software design, power-saving software strategies, resource efficiency, paperless communication, green evaluation, and e-waste management~\cite{salam2016developing}.
    \item Research into the sustainability of software ecosystems remains in its infancy~\cite{da2017has}.
\end{itemize}

Below, we build on these reviews in three main ways: 
\begin{enumerate}
    \item We inductively generate an improved theory of sustainability.
    \item Our scoping review included a larger sample of primary studies.
    \item We employ meta-synthesis (qualitative synthesis of predominately qualitative research).
\end{enumerate}

\section{Method} \label{sec:method}

We reviewed the literature in two phases. First, we conducted a scoping review~\cite{arksey2005scoping} to explore the overall state of research on SSE. This approach helped us determine what sort of systematic review would be most appropriate: meta-analysis, meta-synthesis, case survey or critical review~\cite{ralph2022paving}. For reasons explained below, meta-synthesis (qualitative analysis of predominately qualitative research) was the most appropriate approach, so we conducted a meta-synthesis on a subsample of primary studies to generate an improved theory of SSE. 

\subsection{Sources} \label{sec:sources}

We adopted search-based sampling \cite{baltes2022sampling}. To retrieve the initial set of papers for the review, searches were performed on five databases: \textit{IEEE Xplore}, the \textit{ACM Digital Library}, \textit{ScienceDirect}, \textit{SpringerLink}, and \textit{arXiv}. We selected the first four databases because they are common publication venues for SE research, and the arXiv pre-print server to mitigate publication bias, as explained in Section \ref{sec:implications}.

\subsection{Selection Criteria} \label{sec:inclusion}

The following inclusion criteria were defined: 

\begin{itemize}
    \item The paper was published on or after Jan. 1, 2012
    \item The paper's full text is accessible in English
    \item The paper explicitly relates to sustainability or greenness 
    \item The research context is explicitly SSE or sustainable software products 
\end{itemize}

The following exclusion criteria were defined: 

\begin{itemize}
    \item The paper is only tangentially related to sustainability or greenness
    \item The paper is only tangentially related to software development or software products
    \item The paper is less than 3 pages long
    \item The paper is an introduction to a special issue or workshop
    \item The paper has been withdrawn
\end{itemize}

\subsection{Search String} \label{sec:searchstring}

The search string was conceived to find results related to sustainability or greenness in software or software development. The general search string used is as follows.

\smallskip
{\narrower \noindent \textbf{Search String:} (sustaina* OR green OR ecolog*) AND (``software development'' OR ``software engineering'' OR ``software design'' OR ``software requirement'' OR ``software product'')\par}
\smallskip

\begin{figure}[t]
\centering
%\noindent
\centerline{\includegraphics[width=\columnwidth]{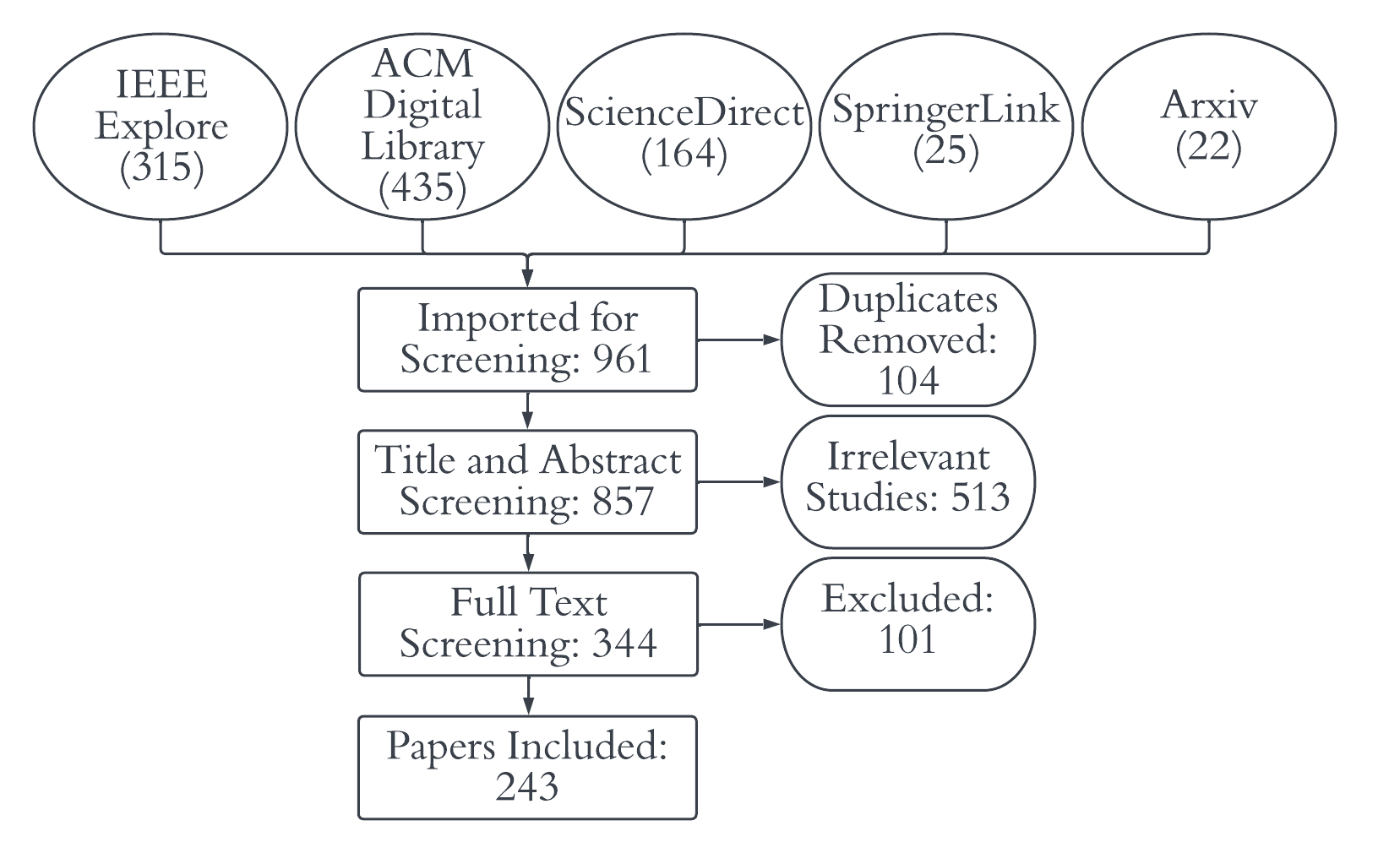}}
%\includesvg[width=\columnwidth]{data_extraction_diagram.png}
\caption{PRISMA diagram~\cite{page2021prisma}}
\label{fig:PSE}
\end{figure}

The search string was modified for each database (see Table \ref{tab:tab_searchStrings}). To improve the relevance of the results, the search string was used on document titles and abstracts where possible instead of full text. Commensurate with the selection criteria, the search results were restricted to papers published on or after January 1, 2012. Figure \ref{fig:PSE} shows the full paper selection process. For ScienceDirect, the search string was used on the title, abstract, and keywords. Lack of wildcard support and limitations on the number of operators allowed in one string led to the query being split into two versions, with “sustaina*” and “ecolog*” replaced by complete words. Due to limitations in SpringerLink’s search engine, the Springer Nature Metadata API was used to run the query. Because abstracts could not be searched without also searching the full text, the query was used to search titles only.

\begin{table*}[t]
\centering
\caption{Search strings used for the review}
\label{tab:tab_searchStrings}
\renewcommand{\arraystretch}{1.3}
\begin{tabularx}{\linewidth}{lX}
\toprule
\textbf{Source} & \textbf{Search String}\\ 
\midrule
IEEE Xplore & (``Document Title'':sustaina* OR ``Document Title'':green OR ``Document Title'':ecolog*) AND (``Document Title'':``software development'' OR ``Document Title'':``software engineering'' OR ``Document Title'':``software design'' OR ``Document Title'':``software requirement'' OR ``Document Title'':``software product'') OR (``Abstract'':sustaina* OR ``Abstract'':green OR ``Abstract'':ecolog*) AND (``Abstract'':``software development'' OR ``Abstract'':``software engineering'' OR ``Abstract'':``software design'' OR ``Abstract'':``software requirement'' OR ``Abstract'':``software product'')  \\
ACM DL & [[[Publication Title: sustaina*] OR [Publication Title: green] OR [Publication Title: ecolog*]] AND [[Publication Title: ``software development''] OR [Publication Title: ``software engineering''] OR [Publication Title: ``software design''] OR [Publication Title: ``software requirement''] OR [Publication Title: ``software product'']]] OR [[[Abstract: sustaina*] OR [Abstract: green] OR [Abstract: ecolog*]] AND [[Abstract: ``software development''] OR [Abstract: ``software engineering''] OR [Abstract: ``software design''] OR [Abstract: ``software requirement''] OR [Abstract: ``software product'']]] AND [Publication Date: (01/01/2012 TO *)]  \\
ScienceDirect (1) & (sustainable OR sustainability OR green) AND (``software development'' OR ``software engineering'' OR ``software design'' OR ``software requirement'' OR ``software product'')  \\
ScienceDirect (2) & (ecological OR ecology) AND (``software development'' OR ``software engineering'' OR ``software design'' OR ``software requirement'' OR ``software product'') \\
SpringerLink & (title:sustainable OR title:sustainability OR title:green OR title:ecological OR title:ecology) AND (title:\textbackslash``software development\textbackslash'' OR title:\textbackslash``software engineering\textbackslash'' OR title:\textbackslash``software design\textbackslash'' OR title:\textbackslash``software requirement\textbackslash'' OR title:\textbackslash``software product\textbackslash'')  \\
arXiv & order: -announced$\_$date$\_$first; size: 50; date$\_$range: from 2012-01-01 ; include$\_$cross$\_$list: True; terms: AND title=sustaina* OR green OR ecolog*; AND title=``software development'' OR ``software engineering'' OR ``software design'' OR ``software requirement'' OR ``software product''; OR abstract=sustaina* OR green OR ecolog*; AND abstract=``software development'' OR ``software engineering'' OR ``software design'' OR ``software requirement'' OR ``software product'' \\
\bottomrule
\end{tabularx}
\end{table*}

\subsection{Paper Collection and Screening} \label{sec:selection}

The query was executed on each database and the full list of references from each was saved. The search process was completed on October 16, 2022 with \totalRetrieved{} results. \numDuplicateManual{} duplicate result was manually removed. The remaining 915 references were uploaded to the Covidence\footnote{\url{https://www.covidence.org/}} review management application, where \numDuplicateAuto{} duplicate results were automatically removed. After duplicate removal, the total number of papers was \numTitleAbScreening{}. 

During the title and abstract screening process, the \numTitleAbScreening{} papers were divided into 17 batches. The first 16 batches contained 50 papers; the final batch contained 57 papers. Each batch was screened independently by two screeners (the first and second authors), who applied the selection criteria by reviewing the abstract and title. After each batch the screeners met with the fourth author to discuss and resolve any disagreements.  %The screeners were two undergraduate students who learned the screening process as they went and were supervised by an experienced researcher who audited the screening. The screeners worked separately until the end of each batch, where they would reconvene and discuss any disagreements. If a consensus could not be reached on whether to include or exclude a paper, the paper was sent to a third researcher to come to a final decision. The experienced researcher acted as the third reviewer and the process for reaching consensus was part of the screeners' training.
As we resolved disagreements, we wrote decision rules to prevent future disagreements; for example:

\begin{itemize}
    \item If sustainability is one of multiple variables considered, \textbf{include} the paper.
    \item If the paper addresses technical sustainability in the context of reuse or influencing a software ecosystem, \textbf{include} the paper.
    \item If software was used to support sustainability in another domain but the paper does not focus on software development or software products, \textbf{exclude} the paper.
    \item If the paper focuses on a process or practice but does not demonstrate its relationship to sustainability, \textbf{exclude} the paper.
\end{itemize}

\begin{figure}[t]
    \centering
    \includegraphics[width=\linewidth]{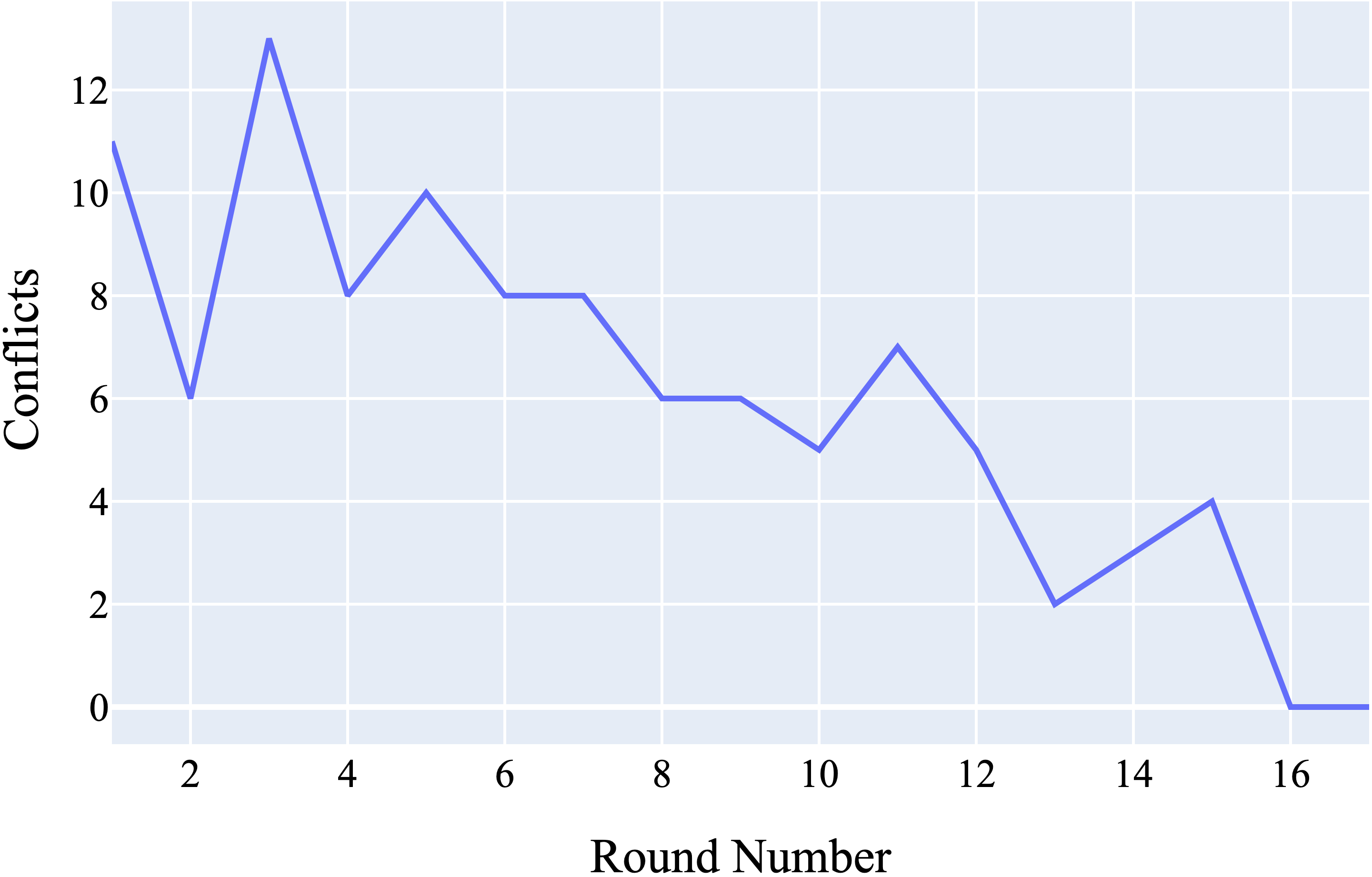}
    \caption{Reviewer Disagreements (Title and Abstract Screening)}
    \label{fig:ReviewerConflictGraph}
\end{figure}

During the screening process, inter-rater agreement (Cohen's Kappa) increased from 0.78 to 1.00, indicating strong to perfect agreement (see Figure \ref{fig:ReviewerConflictGraph}; Table \ref{tab:reliability}). By the end, we reached consensus on all papers and the excluded \numTAScreeningExcluded{}.

\begin{table}[ht]
\setlength{\extrarowheight}{5pt}
\caption{Inter Rater Reliability Table}
\centering
\label{tab:reliability}

\begin{tabular}{crr|crr}
\toprule
\textbf{Round} & \textbf{n} & \textbf{Kappa} & \textbf{Round} & \textbf{n} & \textbf{Kappa} \\
\midrule
1 & 50 & 0.78 & 10 & 50 & 0.9 \\
2 & 50 & 0.88 & 11 & 50 & 0.86 \\
3 & 50 & 0.74 & 12 & 50 & 0.9 \\
4 & 50 & 0.84 & 13 & 50 & 0.96 \\
5 & 50 & 0.8 & 14 & 50 & 0.94 \\
6 & 50 & 0.84 & 15 & 50 & 0.92 \\
7 & 50 & 0.84 & 16 & 50 & 1 \\
8 & 50 & 0.88 & 17 & 57 & 1 \\
9 & 50 & 0.88 & & & \\
\bottomrule
\end{tabular}

\end{table}

The next phase of paper selection was a full text review conducted by the same two screeners. During this phase the screeners divided the remaining papers among them and applied the selection criteria without reconvening. As the screeners' had developed a detailed, shared understanding of what papers should be included, having both screeners independently read and screen the full text of all \numFullTextScreening{} papers seemed unnecessary. Each time a paper was excluded, the screener indicated the reason for exclusion in Covidence. All full text exclusions were reviewed by the fourth author. In total, \numFTScreeningExcluded{} papers were excluded during full text review for the following reasons:  

%considered relevant was clearer. The 319 papers in the second phase were thus not divided into batches. As part of the training process, any paper deemed irrelevant by either of the screeners was sent to the same third reviewer for a final decision, along with a description containing reason(s) why it was thought to be irrelevant. If the experienced researcher disagreed with any decision to exclude a paper, the paper was added back into the pool and kept for the data extraction phase. If the experienced researcher agreed with any decision to exclude, the paper was excluded and the reason for exclusion was noted. By the end of the full text review phase, 90 papers were excluded and \numPrimaryStudies{} papers were deemed relevant. Of the 90 excluded papers, the reasons for exclusion were as follows.

\begin{itemize}
    \item 77 were only tangentially about sustainability
    \item 14 were not about software development or sustainable software products
    \item 4 were less than three pages long 
    \item 3 were introductions to a special issue or workshop 
    \item 2 were not available in English
    \item 1 paper was withdrawn
\end{itemize}

\subsection{Data Extraction} \label{sec:extraction}

Data extraction took place simultaneously with full text review, using a Covidence data extraction form. For each paper, we extracted the following information: list of authors, title, year of publication, publication type (journal, conference or book chapter), topic of study, objective, research method, definition of sustainability used, dimensions of sustainability considered (ecological, social, individual, economic, or technical), and whether the paper focused on software development or software products. Data extraction of all \numPrimaryStudies{} primary studies was completed on November 18, 2022.

Data extraction was mostly straightforward, except for the research method used by each paper. During data extraction, the research team discussed many of the papers, noting a large amount of method slurring (i.e. claiming to use a method one has not actually used~\cite{stol2016grounded}) and more general confusion over the meaning of different methodological terms. The empirical method used in [P161], for example, was initially categorized as an experiment but was changed to engineering research and quantitative simulation due to lack of experimental controls. To label research methods used, we used the empirical method taxonomy and definitions from the Empirical Standards for Software Engineering Research~\cite{ralph2020standards}.

Most of the data cleaning and analysis was conducted in a Jupyter Notebook\footnote{\url{https://jupyter.org/}} (see Section \ref{sec:availability}) using NumPy (v. 1.20.3), Pandas (v. 1.3.4), and Plotly (version 5.6.0).

\subsection{Synthesis Options} \label{sec:synthesis_options}

Based on the research methods used in the collected papers, some synthesis methods were not feasible. For example, there were not enough experiments for a useful meta-analysis or enough case studies for a meaningful case survey.

Given the number of qualitative studies, a qualitative meta-synthesis~\cite{finlayson2008qualitative, nye2016origins} is feasible. This type of synthesis is attractive for gaining insight on perceptions of sustainability as it exists in the real world, given that the included studies consider sustainability in real-world contexts. Another option, given the amount of non-empirical papers, is a critical discourse analysis~\cite{fairclough2013critical}. We selected meta-synthesis for two reasons: \begin{enumerate}
    \item analyzing empirical research would support theory-building better than analyzing non-empirical research;
    \item multiple synthesis methods would be too difficult to present in a single paper.
\end{enumerate}

\subsection{Synthesis Approach} \label{sec:synthesis_approach}

The data synthesis was conducted by the whole research team. The full list of extracted studies was reviewed and all articles that performed case studies, grounded theory, interview studies, or document analysis were included in the synthesis subsample---\numQualitative{} primary studies in total. %During synthesis, six of the papers were found to not match the criteria for a qualitative paper and were recategorized, leaving a final set of \numQualitative{} papers for synthesis.

These \numQualitative{} papers were divided between the first three authors, who summarized them and identified thematic connections between the papers in their respective sets. These connections were identified by reading through each paper and noting down key themes, concepts, or conclusions related to sustainability, such as how sustainability was defined, perceived, or achieved in the paper. After identifying these themes within each paper, each researcher reviewed their notes from previous papers and iteratively identified new similarities within the set as each new paper was read.

After each researcher identified connections in their set of papers, the group reconvened and identified connections between the entire set of papers. This took place as a group discussion where each researcher presented their notes to the other researchers and integrated them by comparing them to the connections identified by the other researchers. While comparing each researchers' respective notes, connections that appeared within the entire pool were identified and discussed. After identifying several shared connections between all the papers, the group held several collaborative drawing sessions to visualize these connections in a model. The group iterated through various drawings and eventually developed the novel model of SSE presented below.

\section{Analysis and Results} \label{sec:results}

\begin{figure*}[t]
    \centering
    \includegraphics[width=\linewidth]{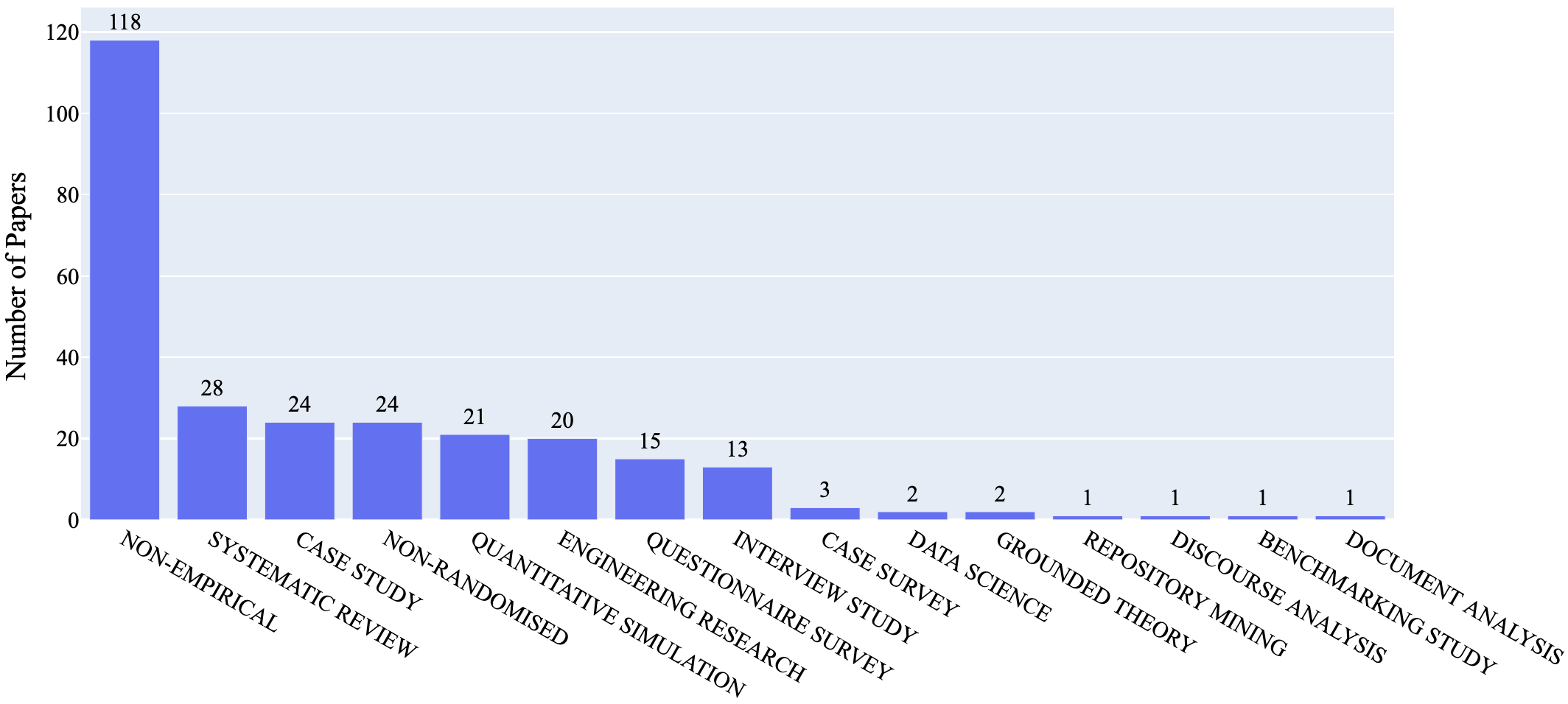}
    \caption{Research Approaches used by the primary studies}
    \label{fig:fig_methodsUsed}
\end{figure*}

\subsection{Empirical Methods used by Primary Studies} \label{sec:results_method}

Of the \numPrimaryStudies{} papers selected for the data extraction process, 118 (49\%) did not include an empirical study. For the empirical studies, the most frequently used methods were systematic review, case study, non-randomized technical experiments (i.e., experiments without random assignment) and quantitative simulations (i.e., simulations or experiments that assess technological artifacts using using mathematical models)~\cite{ralph2020standards}, (see Figure \ref{fig:fig_methodsUsed}). The sum of all the method usages (each method type in Figure \ref{fig:fig_methodsUsed}) is greater than the total number of papers selected (\numPrimaryStudies{}) because some papers use multiple empirical methods and therefore appear in multiple categories. For example, study [P18] is both engineering research and a quantitative simulation. 

\begin{figure}[ht!]
    \centering
    \includegraphics[width=0.8\linewidth]{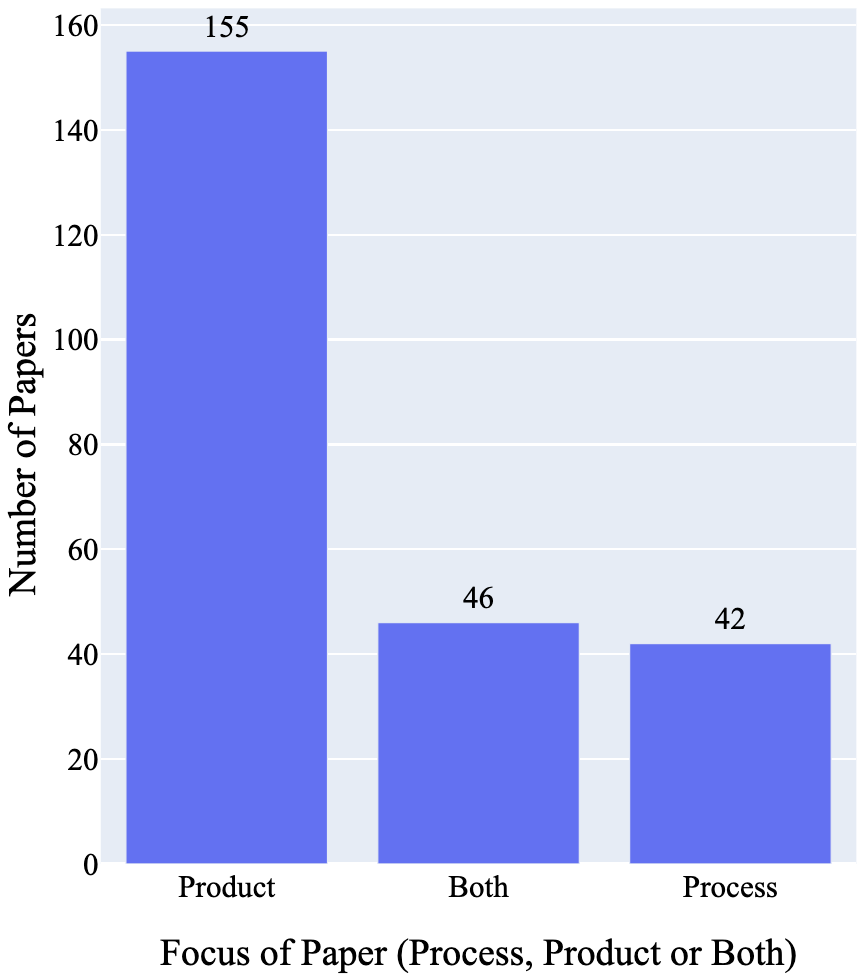}
    \caption{Count of papers considering each focus category}
    \label{fig:fig_focusCategory}
\end{figure}

\subsection{Focus of Primary Studies (Process or Product)} \label{sec:results_focus}
Discussions around sustainability in SE usually focus on processes or products. For example, development activities such as business trips may increase the ecological footprint of a software company~\cite{Markus2010Green} while a software product may consume high amounts of energy during use~\cite{debbarma2016green}. Here, \textit{process} refers to the activities that make up software development, while \textit{product} refers to a software artifact that is created by a software development process. Papers may also investigate both processes and products (e.g. [P170]). In our analysis, 155 papers (64\%) addressed only the sustainability of software products, 42 (17\%) focused exclusively on software development processes, and 46 (19\%) examined both (see Figure \ref{fig:fig_focusCategory}).

\begin{figure}[ht!]
    \centering
    \includegraphics[width=0.8\linewidth]{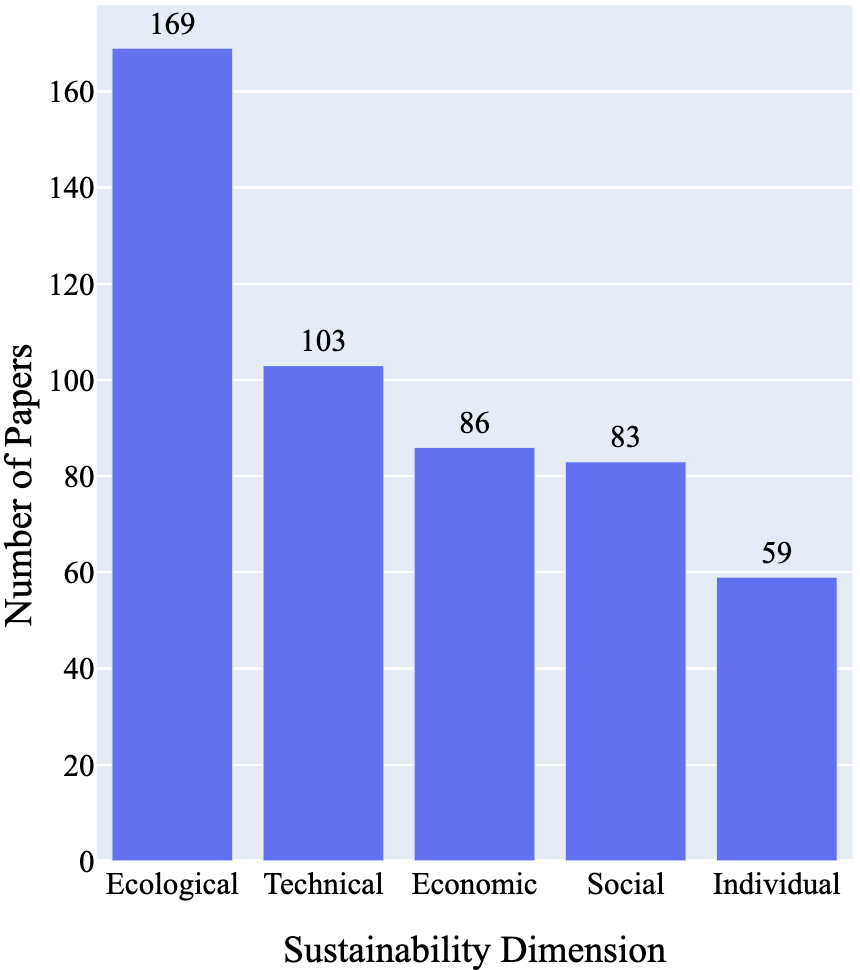}
    \caption{Count of papers considering each sustainability dimension}
    \label{fig:fig_sustainabilityDimensions}
\end{figure}

\subsection{Sustainability Dimensions Considered} \label{sec:results_dimensions}
Figure \ref{fig:fig_sustainabilityDimensions} shows the distribution of papers by sustainability dimension. Sustainability dimensions are \textit{not} mutually exclusive. For example, [P117] considered both the economic and technical dimensions of sustainability. The most commonly considered dimension is ecological sustainability, which is addressed by 169 (70\%) of the \numPrimaryStudies{} papers, followed by technical (42\%), economic (35\%), social (34\%), and individual (24\%). 145 (60\%) of the \numPrimaryStudies{} papers consider only one dimension of sustainability e.g., P79 considers only ecological sustainability while P95 considers only economic sustainability. The most common sustainability dimension to be considered by itself was ecological (93), followed by technical (37), social (6), individual (5), and economic (4). All dimensions were considered by 35 studies, while 19 studies considered at least four, and 16 studies considered at least three.

\begin{table}[t]
\centering
\caption{Most prolific authors.}
\label{tab:authorFrequency}
\begin{tabularx}{\linewidth}{lXr}
\toprule
\textbf{Author} & \textbf{Papers} & \textbf{Count} \\ 
\midrule
Birgit Penzenstadler & P223, P216, P215, P212, P211, P208, P196, P183, P182, P163, P132, P114, P77, P74, P63, P48, P21, P17, P16, P232 & 20 \\
Coral Calero & P182, P154, P147, P115, P94, P93, P87, P21, P11, P9, P236 & 11 \\
Leticia Duboc & P223, P222, P215, P208, P196, P183, P74, P68, P232 & 9 \\
Monica Pinto & P145, P112, P106, P96, P78, P22, P2, P1 & 9 \\
Lidia Fuentes & P145, P120, P112, P106, P96, P78, P22, P2, P1 & 9 \\
Patricia Lago & P224, P178, P151, P128, P111, P110, P103, P27 & 8 \\
Stefanie Betz & P223, P208, P198, P196, P183, P114, P74, P232 & 8 \\
Ruzanna Chitchyan & P223, P215, P208, P196, P183, P166, P74, P53 & 8 \\
Colin Venters & P223, P215, P208, P196, P183, P114, P74, P232 & 8 \\
Debra Richardson & P218, P216, P212, P211, P209, P48, P21, P8 & 8 \\
Norbert Seyff & P223, P215, P208, P196, P183, P74, P232 & 7 \\
Christoph Becker & P223, P215, P208, P196, P183, P114, P74 & 7 \\
Eva Kern & P189, P105, P102, P11, P7, P5, P231 & 7 \\
Stefan Naumann & P189, P105, P102, P11, P7, P5, P231 & 7 \\
Mario Piattini & P221, P133, P115, P87, P73, P9 & 6 \\
Felix Garcia & P154, P115, P94, P93, P11 & 5 \\
Maria Angeles Moraga & P154, P147, P115, P93, P236 & 5 \\
Daniel-Jesus Munoz & P145, P120, P112, P28, P22 & 5 \\
\bottomrule
\end{tabularx}
\end{table}

\subsection{Most Prolific Authors} \label{sec:results_authors}
\begin{figure}[t]
    \centering
    \includegraphics[width=\linewidth]{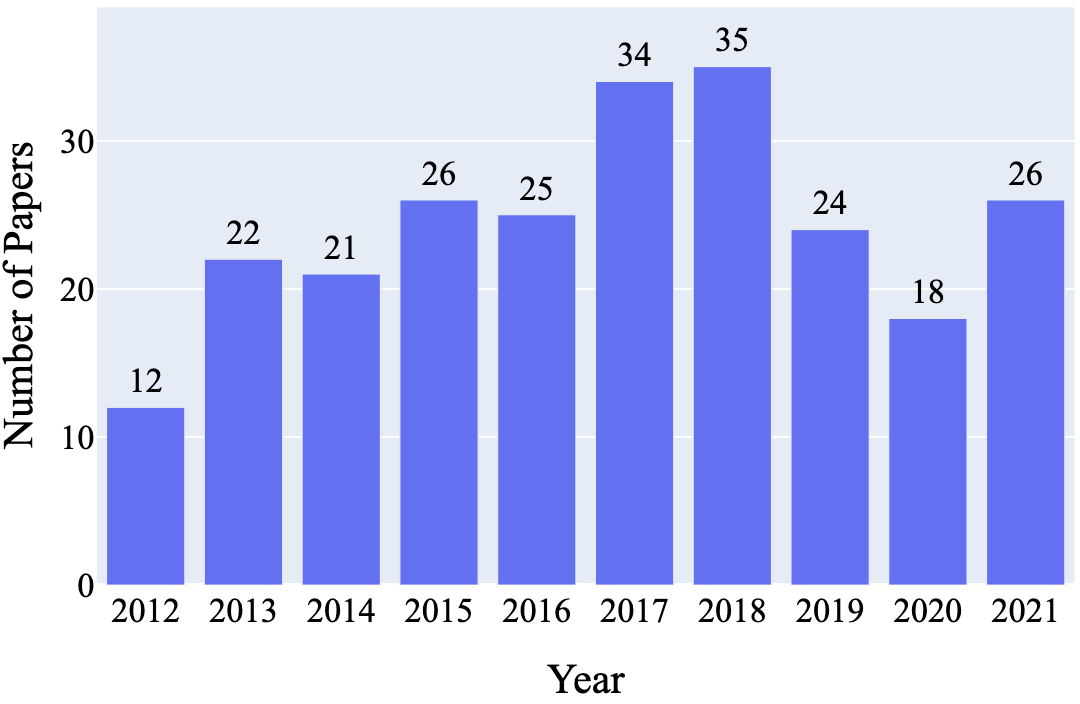}
    \caption{Count of papers published in each year since 2012}
    \label{fig:fig_yearsPublished}
\end{figure}

There were several authors who had multiple papers in our data set. In total, 96 authors appeared more than once in our data set. Of those 96 authors, 18 appeared at least five times (see Table \ref{tab:authorFrequency}). The number of publications per year peaked in 2018 (see Figure \ref{fig:fig_yearsPublished}).

%\subsection{Temporal Trends} \label{sec:results_years}

\section{Synthesis} \label{sec:synthesis_results}

We aggregated the results of the subsample of \numQualitative{} qualitative studies as explained in Section \ref{sec:synthesis_approach}. Through this process, we synthesized six key propositions, and a new model for SSE based on these propositions, as follows. Our most significant departure from existing literature on sustainable development is the proposition that \textit{sustainability is stratified}, discussed next. 

\subsection{Sustainability is Stratified} \label{sec:stratified}

Prior research typically conceptualizes sustainable development in terms of three ``intersecting'' dimensions or ``pillars'': economic, environmental, and social \cite{purvis2019three}. We observe that each dimension appears stratified. Stratified simply means `comprising multiple layers'. Sustainability has different meanings at different strata, and a project that appears sustainable at one stratum may seem unsustainable at another.

Economic sustainability could refer to the economy of the whole world or a specific country, region, municipality, industry, organization, or individual [P149, P206, P220]. Consider, for example, a company that generates novels using a large language model and charges readers a flat monthly fee to read as much as they want. At the \textit{organization layer}, economic sustainability refers to the product's capacity to produce enough revenue to pay off the cost of building it, cover ongoing maintenance, and return a reasonable profit. At the \textit{individual layer}, economic sustainability refers to the extent to which readers can afford the subscription fee. At the \textit{industry layer}, economic sustainability refers to the product's effects on the profitability of conventional publishing (e.g., computer-generated novels become so good that publishing human-generated novels becomes economically nonviable). At the \textit{national layer}, economic sustainability refers to the product's effects on a country's economy (e.g. gross domestic product, unemployment rates, inflation). While novel publishing may be too small to perturb a national economy, other software systems can have significant nation-level effects (e.g. Bitcoin undermines individual nations' control over their money supply).

Meanwhile, the natural environment can refer to anything from the entire world to specific patch of bacteria, depending on our context. The ecological sustainability of software is often equated with energy use [P62, P94, P151], which makes sense for our computer-generated novel example where training large language models requires much electricity. However, software products have other complex effects on the natural environment [P216, P220]. Software runs power grids, mining equipment, environmental monitoring stations, agricultural irrigation, climate modeling, and sewage treatment plants. Such systems can have differential effects at different layers. For example, an agricultural information system could simultaneously benefit biodiversity within a farm (e.g. by supporting native ground cover) while harming the larger nearby river system through pesticide runoff. 

Similarly, social sustainability can refer to global, national or regional societies [P117, P126], communities [P116], organizations [P23, P86], or the social implications of a system for an individual person [P59]. The Karlskrona manifesto's~\cite{becker2015sustainability} addition of the ``individual'' dimension attempts to capture part of this stratification. Social sustainability may refer to a system's impact on: the individual's physical or psycho-social wellbeing (\textit{individual level}); team cohesion and resilience (\textit{team level}); organizational culture (\textit{organization level}); or pro-social vs. anti-social affordances (\textit{societal level}). 
 
Likewise, technical sustainability concerns range from code smells [P72, P85] and low-level design decisions [P172, P186, P206], to whole software artifacts [P60], to entire software ecosystems [P34, P58, P91]. Although these concerns exist at different scales, they all impact technical sustainability. 
 
The point here is not simply that software products can have different effects at different levels. The point is that sustainability has different \textit{meanings} at different levels. Nuanced assessment of a system's sustainability therefore demands examining its effects layer by layer.

\subsection{Sustainability is Multisystemic} \label{sec:multisystemic}

A phenomenon is \textit{multisystemic} when it emerges from the interactions among several systems. A person's resilience (capacity to overcome challenges), for example, emerges from their neurophysiology, physical environment, and socio-cultural context~\cite{ungar2020resilience}.

Similarly, consider the sustainability of a mobile app for investing. Sustainable development is about meeting today's needs without sabotaging tomorrow's needs. Suppose the investing app seeks to meet today's need to wisely invest our retirement savings. The sustainability of this app emerges from the intersection of several systems:
\begin{itemize}
    \item The user's neurophysiological system. If the app presents information in a manner that is inaccessible or confusing [P59], we may make bad investment decisions, undermining our ability to retire comfortably in the future. 
    \item The national banking system. Apps may be subject to government regulations [P212, P220], which may make the app more sustainable in some ways (e.g., by mandating disclosure of important information) but less sustainable in others (e.g. by excluding small investors from the most profitable opportunities).
    \item Our collective technical infrastructure. If the app is developed using unstable technologies, and embedded in an unstable software ecosystem [P43, P180], our retirement savings may be at risk.
    \item The investment vehicles themselves. Investing in hydroponics companies and solar panel manufacturers is more sustainable than investing in oil companies and bitcoin.
    \item The digitization and automation of the IT services can improve scalability and efficiency while simultaneously lead to job and role losses  (e.g. accountants) [P243].
\end{itemize}
    
The primary studies do not refer to sustainability as multisystemic per se. Many primary studies describe different ``pillars'' or ``dimensions'' of sustainability---``environmental'', ``economic'', ``social'', etc. However, when we aggregate and critically reflect on these studies, conceptualizing sustainability as multisystemic has three advantages:
\begin{enumerate}
    \item The systems from which sustainability emerges vary. In the above example, it was more informative to think about the neurophysiological, banking, and infrastructural systems than ``economic'', ``environmental'' and ``social'' dimensions. 
    \item Describing sustainability in terms of dimensions emphasizes the disconnect between them: a product can be economically sustainable but environmentally unsustainable or environmentally sustainable but socially unsustainable. Describing sustainability as multisystemic emphasizes the intersections among the systems: a product's (or process's) sustainability results from human physiological, psychological, societal and economic systems colliding with each other and with ecological systems. 
    \item Thinking of sustainability as multisystemic, and recognizing that many of these constituent systems are unpredictable and responsive, suggests applying a complex adaptive systems perspective~\cite{holland1992complex} for analyzing sustainability.
\end{enumerate}

\subsection{Process Sustainability differs from Product Sustainability} \label{sec:processDiffers}

SE impacts sustainability through both the process of development and the software products generated~\cite{penzenstadler2013generic}. SSE processes can promote sustainability by mitigating disruption during development [P23], motivating development communities [P34, P58, P118, P125], and sustaining business relationships [P149]. Sustainable software products can promote sustainability through reduced energy consumption [P62, P85, P151], software quality [P72, P149, P180], pro-social impact [P126], and technical durability [P60]. 

Both process and product may have mixed sustainability profiles. For example, a company can simultaneously implement policies that are environmentally sustainable (e.g. encouraging employees to take active transport) and socially unsustainable (e.g. encouraging a culture of overwork). Similarly, a product that generates lots of revenue while using huge amounts of energy may be economically sustainable but ecologically unsustainable. Likewise, a sustainable process could produce an unsustainable product and an unsustainable process could produce a sustainable product.

\subsection{Processes Determines Products} \label{sec:processDeterminesProduct}

Software products do not appear out of thin air; they are built by developers via a software development process. Development activities such as user-centered design [P59] and stakeholder modeling [P216, P212], for example, are used to develop software products that reflect human concerns, while the unified modeling language [P72] and pair programming [P23] are used to develop products with high-quality code. The processes used during software development will influence the software product that is ultimately produced.

\subsection{Technical Sustainability is a Product (not Process) Attribute} \label{sec:technicalSustainability}

Technical sustainability ``refers to longevity of information, systems, and infrastructure and their adequate evolution with changing surrounding conditions [including] maintenance, innovation, obsolescence, [and] data integrity'' \cite[p. 471]{becker2015sustainability}. Technical sustainability therefore applies to software products, not software development processes. It relates to product aspects such as code quality [P23, P60, P72, P180] and longevity, especially in open-source infrastructure [P58, P91, P125]. The development process may \textit{cause} the technical sustainability of the product (e.g. by limiting technical debt), but technical sustainability is an \textit{attribute} of the product.

Like the other dimensions, technical sustainability is stratified and multisystemic. It can refer to the robustness of anything from a piece of code [P72] to the collective technological infrastructure of humanity [P43]. The sustainability of any particular technological artifact emerges from the intersection of many subsystems such as the programming language(s) in which it is written, the third-party libraries it uses, and the database or other technology used to store data.

\subsection{Sustainability is Participatory} \label{sec:participatory}

Achieving sustainability is a participatory process. \textit{Participatory} means that achieving sustainability requires support from many involved parties, not isolated efforts. It involves identifying and consulting stakeholders for sustainability, as stakeholder concerns impact software development and certain dimensions of sustainability may be neglected without dedicated advocates [P216, P220]. Sustainability requires participation from diverse user groups, including those with less technological experience and differing support needs, to both develop sustainably and create sustainable products [P59, P126]. It may be considered a matter of human values, such as trust [P126, P216], which reflect the needs of humanity more broadly. Several studies found that software practitioners also view sustainability as a shared responsibility rather than an individual endeavor [P62, P180, P186].

\section{The Stratified Model of Software Sustainability} \label{sec:synthesis_model}

\begin{figure*}[t]
    \centering
    \includegraphics[width=0.7\linewidth]{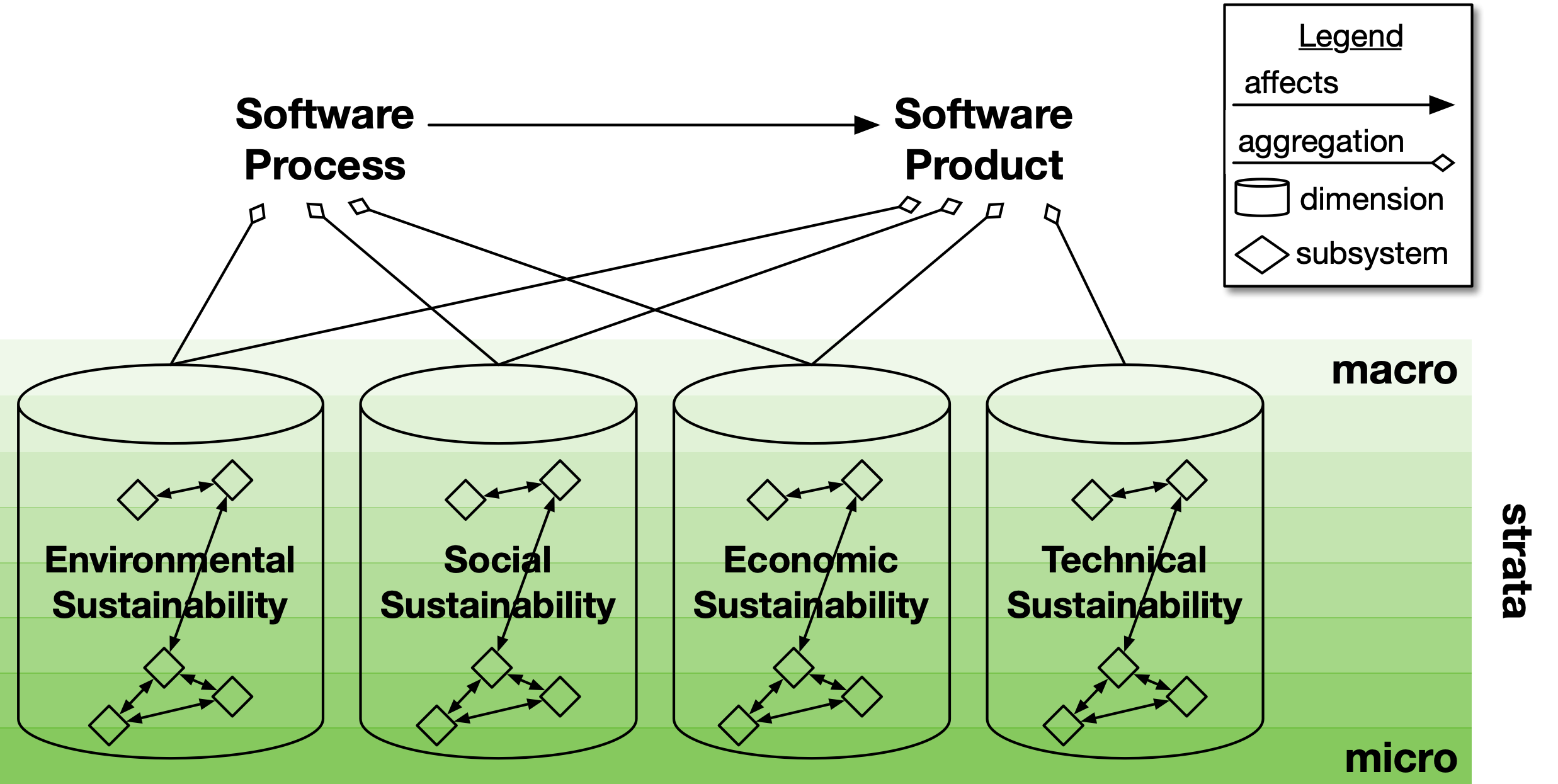}
    \caption{The Stratified Theory of Sustainable Software Engineering}
    \label{fig:fig_sustainabilityModel}
\end{figure*}

Based on the propositions elucidated in the preceding section, we posit a novel model of Sustainable Software Engineering (Figure \ref{fig:fig_sustainabilityModel}). The goal of this model is to explain how sustainability relates to software development based on existing qualitative research and to provide a more nuanced view of the stratified and multisystemic nature of sustainability according to our observations from the meta-synthesis. The previous section \textit{justifies} our propositions; this section \textit{explains how the model illustrates} our propositions.

First, the model separates the software process from the software product to illustrate that (1) sustainability is a property of both (Section \ref{sec:processDiffers}), and (2) while the nature of the software development process affects the resulting software system, a sustainable process does not guarantee a sustainable product or vice versa (Section \ref{sec:processDeterminesProduct}). 

Second, the model depicts four dimensions or pillars of sustainability---environmental, social, economic and technical. All four dimensions apply to software products. Technical sustainability does not apply to the software process (see Section \ref{sec:technicalSustainability}). The model does not include an ``individual'' pillar because the individual is the just the first strata of the social pillar.  

To illustrate the stratified nature of sustainability (see Section \ref{sec:stratified}), the four pillars rest on a color gradient. This gradient represents a series of layers from micro (e.g. an individual person, organism, or function) to macro (e.g. the global environment, society, economy, or humanity's cumulative technical infrastructure). Precisely which and how many strata are relevant to any particular sustainability analysis depends on the context. The point being illustrated is that each pillar entails numerous layers and sustainability has different meanings at different layers.

To illustrate the multisystemic nature of sustainability (Section \ref{sec:multisystemic}), each pillar contains a constellation of social, technical, or sociotechnical subsystems. Again, the specific number of and interconnections between these subsystems is context-dependent. The point being illustrated is that even within one dimension, sustainability may emerge from the intersection of two or more systems. Indeed, even within one strata of one dimension, sustainability may emerge from multiple interacting systems. For example, the social sustainability of a dating application may emerge from the intersection of multiple user groups, all at the same \textit{community layer}. 

In sum, the model illustrates the inherent complexity of either assessing sustainability in the context of SE, or approaching sustainability as a participatory process (Section \ref{sec:participatory}).

\section{Discussion} \label{sec:discussion}

The proposed theory of SSE improves on Becker et al.'s~\cite{becker2015sustainability} 5-pillar model in three main ways: 

\begin{enumerate}
    \item It emphasizes the stratified nature of sustainability; drawing a more nuanced picture of the diverse meanings of sustainability across strata. Explicating this stratification allows us to combine Becker et al.'s \textit{individual} and \textit{social} dimensions. 
    \item It emphasizes the multisystemic nature of sustainability; namely, that even within a dimension / strata, sustainability may emerge from the intersection of several social, technical, or sociotechnical subsystems.
    \item It distinguishes the sustainability of software processes from the sustainability of the resulting software products.
\end{enumerate}

Stratifying sustainability means thinking about how sustainability manifests at different levels of abstraction (e.g., a local ecosystem vs. the global ecosystem; individual vs. community impacts; organizational profits vs. gross domestic product). Because individual and social sustainability both deal the human wellbeing, but at differing scales~\cite{becker2015sustainability}, they are both reflected in the social dimension of the proposed theory. 

Our stratification relates to but differs from Hilty and Aebischer's \cite{hilty2015ict} three layers of impact: ``lifecycle'' (effects of physically producing, maintaining, recycling and disposing ICT hardware), ``enabling'' (production or consumption enabled by applying ICT), and `structural' (changes in economic or institutional structures). Fundamentally, Hilty and Aebischer classify impacts into levels whereas we propose that sustainability means different things at different strata.  

Referring to sustainability as \textit{multisystemic} infuses greater diversity into our conceptualization of SSE. For some software products (e.g. a web-based video game), the environmental dimension may indeed reduce to carbon footprint. However, imagine a software system controlling futuristic efforts to engineer the climate (e.g. by releasing/collecting reflective particles into/from the upper atmosphere). Such a system could have wildly different effects on forests, deserts, wetlands, jungles, plains, mountains, lakes and oceans.

Conceptualizing sustainability dimensions as multisystemic and stratified dovetails with another key finding of our synthesis: that sustainability is a participatory process. \textit{Participatory} means that involving many stakeholders helps with improving sustainability (because the stakeholders bring different perspectives, concepts, ideas and values that would not otherwise be obvious or seem important to the developers). Breaking down a sustainability dimension into its constituent systems and layers should facilitate identifying stakeholders that may not have been obvious from considering the dimension as a single monolothic concern.
By visualizing the systems that impact sustainability, the model also provides a way to visualize the stakeholders within those systems and emphasize with their respective roles in sustainability efforts.

Meanwhile, the proposed theory also differs from prior research in its separation of software processes from the products of those processes. Including product and process in the model acknowledges that sustainability in software is not solely about producing sustainable products but also about developing these products in a sustainable way, such that the entire life of a software artifact is sustainable. Differentiating between process and product aspects of software development is significant not only because of the relative dearth of research on SSE processes compared to sustainable software products but also because the process of developing some software systems can have enormous sustainability challenges. For example, training large language models is highly carbon-intensive \cite{patterson2021carbon} while video game development is plagued by socially unsustainable cycles of crunch and burnout \cite{edholm2017crunch}. 

Becker et al.'s \cite{becker2015sustainability} addition of the technical pillar was crucial for understanding SSE because of the important relationship between technical features and sustainability in software engineering. The proposed model builds on Becker et al.'s insight by clarifying that technical sustainability is (1) an attribute of a product, not a process; and (2) stratified like the other pillars. Technical sustainability entails different meanings at different levels of abstraction, from ``will this function be easy to modify in the future?'' to ``how does this product impact the technical infrastructure of humanity?''

\subsection{Implications} \label{sec:implications}

The proposed model has important implications for both researchers assessing sustainability and practitioners approaching sustainability as a participatory process. To assess sustainability, the model implies many relevant questions:
\begin{itemize}
    \item Are we assessing the software product or its development process?
    \item Which dimension(s) of sustainability are we assessing---environmental, social, economic or technical?
    \item For each dimension we are assessing, what strata are relevant? What does sustainability mean at each stratum?
    \item What are the pertinent subsystems comprising each dimension we are assessing? What does sustainability mean from the perspective of each subsystem? 
    \item How does the software product or process affect each identified subsystem? 
\end{itemize}

Meanwhile, to approach sustainability as a participatory process, the model implies asking which dimensions, layers and subsystems the participants should come from, and whether they are participating in the design of the process or product? In other words, the proposed model is intended to discourage researchers and professionals from simply mapping sustainability concerns to dimensions, and encourage more nuanced thinking about how sustainability manifests in different systems at different levels. Ideally, it pushes us to consider what \textit{kind} of sustainability we mean. 

Our scoping review also has important implications for the SE research community. The most common research method we found was no research method at all. This raises concerns about the emphasis on speculation over evaluation when studying sustainability. Moreover, we found \textit{zero} controlled experiments with human participants that assessed the effectiveness of sustainability interventions. While research on SSE may still be immature, more research that proposes and evaluates sustainability interventions is plainly needed. Interventions can be evaluated in many ways including action research, lab-based experiments with human participants, and field experiments~\cite{ralph2020standards}.  

For software professionals, the absence of controlled evaluations of sustainability interventions means that all sustainable development practices should be viewed with suspicion. Is it really a ``sustainable software development practice'' if there is no evidence that applying the practice increases sustainability? Software educators who teach sustainable development practices should be similarly wary.

We also found that the sustainability discourse in software development centers on software products rather than software development processes. This is unfortunate because the process may be the main source of problems. For example, social sustainability includes the effects of a project not only on society but also on the software development team, and many teams suffer from their organizations' toxic development cultures. Similarly, the environmental impacts of some projects are predominately in the development process rather than the operation of the product (e.g. training the GPT-3 language model~\cite{patterson2021carbon}). More attention to the sustainability of software development processes is, therefore, sorely needed.

Consistent with previous reviews, we found that social and economic sustainability are studied less than ecological and technical sustainability, with economic, social, and individual sustainability studied almost exclusively alongside other dimensions and rarely as the singular focus of a paper. This disparity may imply a de facto hierarchy of importance in SSE research, which conflicts with the view that all dimensions of sustainability are equally important~\cite{becker2015sustainability}. This is especially concerning since so many software products fail or incur extra costs~\cite{nelson2007project}, and because software professionals are at risk of unsustainable behavior such as burnout~\cite{sonnentag1994stressor}.  More research on SSE processes is needed if professionals wish to achieve sustainability throughout software development and not just in the usage and maintenance of software products. SE educators should consider ways of better addressing sustainability in general, and social and economic sustainability in particular, in their courses. Because typical SE curricula are so full already, we suggest integrating sustainability into existing topics and projects, rather than creating whole new modules or courses. Software professionals must also be encouraged to view sustainability not just as an ecological concept or a technical requirement, but also as a concern of human and economic stakeholders.

\subsection{Limitations} \label{sec:limitations}

We used keyword searches to improve replicability, but these searches may miss some relevant papers, introducing sampling bias. Publication bias is also a concern; we attempted to mitigated this threat by including arXiv preprints~\cite{ralph2020standards}, but not all unpublished manuscripts are deposited on arXiv. We optimized reliability by have two independent screeners assess each title and abstract for inclusion, resolving disagreements by discussion (with a third, more experienced researcher), and documenting our decision rules, in a series of discrete rounds. However, having two analysts extract data independently did not seem warranted due to the straightforward nature of the data; this may have harmed reliability. Similarly, including diverse \textit{quantitative} studies in a \textit{qualitative} synthesis was intractable at this stage, which may have affected the resulting propositions and theory in ways we cannot guess. The proposed theory itself is more nuanced than previous attempts, but also more complicated. We believe it represents a good middle ground between monolithic dimensions (too simple) and enumerating all possible strata and subsystems (too complicated). However, a different balance may be optimal in some situations. 

\subsection{Future Work} \label{sec:future}
We plan to \textit{propose and empirically evaluate specific sustainability interventions} and encourage other researchers to do the same. Since countless interventions could potentially improve the sustainability of software processes and products, many researchers can contribute to this area in parallel. 

Interventions may be social (e.g. do more demographically diverse teams produce more socially sustainable software products?), technical (e.g. CPU optimization), or sociotechnical (e.g. does adding energy consumption tests to test-driven development lead to more sustainable software design?). Interventions may focus on software products (e.g. how do we reduce the carbon footprint of hot fixes?) or software processes (e.g. does pair programming make software teams more resilient?). Interventions may focus on technical sustainability (is object-oriented code with higher cohesion and looser coupling more maintainable?), social sustainability (e.g. do undergraduate ethics courses reduce the tendency to design anti-social algorithms?), economic sustainability (e.g. how do we design software more people will actually \textit{want} instead of tricking people into buying software they will hate), or combinations thereof (e.g. how can video game micro-transaction systems be profitable without promoting gaming compulsion?). And none of the above are the research questions or interventions our team plans to investigate next.

Furthermore, more collaboration between researchers and professionals is needed both to develop and to evaluate sustainability interventions that are useful and usable in real-world software development settings. Such collaborations require a dual willingness of software organizations to participate in research and researchers to be receptive to real world problems.

\section{Conclusion} \label{sec:conclusion}

This study explores existing research on sustainability in SE. Our synthesis of qualitative research found that sustainability is multisystemic, stratified, and applies to both software products and the processes that create them. We proposed a theory of sustainable software engineering (Fig. \ref{fig:fig_sustainabilityModel}) that extends prior work (e.g. \cite{hilty2015ict,becker2015sustainability}) by emphasizing the subsystems and layers within each dimension of sustainability.   

Meanwhile, our analysis discovered \textit{zero} controlled experiments; indeed, the dominant research method was non-empirical (e.g. position papers). While non-empirical scholarship can make important contributions, the ratio of essays to empirical studies and the lack of rigorous assessment of sustainability-improving interventions is concerning. We also found that research focuses heavily on the sustainability of software products (vs. development processes) and ecological (vs. economic and social) sustainability. %These gaps imply that common understandings of sustainability in software development are largely speculative, product-focused, and ecologically-minded.

In conclusion, SE research should (1) propose and rigorously assess more sustainability-improving interventions, and (2) develop more sophisticated instruments that account for the different meanings of sustainability at different strata and the diverse social, technical, and sociotechnical systems affected. 

%In our synthesis, we found that sustainability is a multisystemic, stratified, and participatory issue, which is not addressed in the traditional 5-pillar model of software sustainability~\cite{becker2015sustainability}. We proposed a new model for sustainability software development that stratifies sustainability, making the systems and stakeholders that interact with sustainability at different levels of reality more clear. The proposed model can be used to guide future research on SSE. Filling the identified gaps in software sustainability literature will help researchers develop a more comprehensive view of sustainability in software development and support the development of evidence-based interventions that can be performed to improve sustainability in the software industry.

\section*{Acknowledgments}
This project was supported by NSERC Discovery Grant RGPIN-2020-05001, Discovery Accelerator Supplement RGPAS-2020-00081 and the NSERC Undergraduate Student Research Award program.

\section*{Data Availability} \label{sec:availability}

A comprehensive replication package is available on \textit{figshare} at \url{https://doi.org/10.6084/m9.figshare.20731855.v1}. It includes the complete list of primary studies, our data extraction spreadsheet and a Jupyter notebook containing our complete analysis scripts. Below, we list the primary studies included in our qualitative synthesis. (Numbers are not sequential because they come from the scoping review step).

\section*{Primary Studies included in Qualitative Synthesis}

\begin{footnotesize}
\begin{description}[leftmargin=!,labelwidth=\widthof{{[P000]}},font=\normalfont]
  \item [{[P23]}] T. Sedano, P. Ralph, and C. Péraire. ``Sustainable software development through overlapping pair rotation,'' in \textit{Proc. of ESEM}. ACM, 2016, pp. 1-10. 
  
  \item [{[P34]}] C. Steglich, S. Marczak, C. R. De Souza, L. P. Guerra, L. H. Mosmann, F. Figueira Filho, and M. Perin. ``Social aspects and how they influence MSECO developers,'' in \textit{Proc. of CHASE}. IEEE, 2019, pp. 99-106.

  \item [{[P43]}] M. Valiev, B. Vasilescu, and J. Herbsleb. ``Ecosystem-level determinants of sustained activity in open-source projects: A case study of the PyPI ecosystem,'' in \textit{Proc. of ESEC/FSE}. ACM, 2018, pp. 644-655.

  \item [{[P56]}] A. Alami. ``The sustainability of quality in free and open source software,'' in \textit{ICSE Companion Proc.}. IEEE, 2020, pp. 222-225.

  \item [{[P58]}] C. R. de Souza, F. Figueira Filho, M. Miranda, R. P. Ferreira, C. Treude, and L. Singer. ``The social side of software platform ecosystems,'' in \textit{Proc. of the 2016 CHI Conf. on Human Factors in Computing Systems}. ACM, 2016, pp. 3204-3214.

  \item [{[P59]}] R. Bekele, I. Groher, J. Sametinger, T. Biru, C. Floyd, G. Pomberger, and P. Oppelt. ``User-centered design in developing countries: a case study of a sustainable intercultural healthcare platform in Ethiopia,'' in \textit{2019 IEEE/ACM Symposium on Soft. Eng. in Africa (SEIA)}. IEEE, 2019, pp. 11-15.

  \item [{[P60]}] M. R. de Souza, R. Haines, M. Vigo, and C. Jay. ``What makes research software sustainable? An interview study with research software engineers,'' in \textit{Proc. of CHASE}. IEEE, 2019, pp, 135-138.

  \item [{[P62]}] Z. Ournani, R. Rouvoy, P. Rust, and J. Penhoat. ``On reducing the energy consumption of software: From hurdles to requirements,'' in \textit{Proc. of ESEM}. ACM, 2020, pp. 1-12.

  \item [{[P72]}] O. Baddreddin and K. Rahad. ``The impact of design and UML modeling on codebase quality and sustainability,'' in \textit{Proc. of CSSE}. ACM, 2018, pp. 236-244.

  \item [{[P85]}] F. Palomba, D. Di Nucci, A. Panichella, A. Zaidman, and A. De Lucia. ``On the impact of code smells on the energy consumption of mobile applications,'' \textit{Information and Software Technology}, vol. 105, pp. 43-55. 2019.

  \item [{[P86]}] P. Gregory, L. Barroca, H. Sharp, A. Deshpande, and K. Taylor. ``The challenges that challenge: Engaging with agile practitioners’ concerns,'' \textit{Information and Software Technology}, vol. 77, pp. 92-104, 2016.

  \item [{[P91]}] Y. Dittrich. ``Software engineering beyond the project-Sustaining software ecosystems,'' \textit{Information and Software Technology}, vol. 56, no. 11, pp. 1436-1456, 2014.

  \item [{[P94]}] J. A. Garcia-Berna, J. L. Fernandez-Aleman, J. M. C. de Gea, A. Toval, J. Mancebo, C. Calero, and F. Garcia. ``Energy efficiency in software: A case study on sustainability in personal health records,'' \textit{J. of Cleaner Production}, vol. 282, pp. 124262, 2021.

  \item [{[P101]}] S. Butler, J. Gamalielsson, B. Lundell, C. Brax, A. Mattsson, T. Gustavsson, J. Feist, and E. Lönroth. ``Maintaining interoperability in open source software: A case study of the Apache PDFBox project,'' \textit{J. of Systems and Software}, vol. 159, pp. 110452, 2020.

  \item [{[P107]}] M. Liu, S. Hansen, and Q. Tu. ``Sustaining collaborative software development through strategic consortium,'' \textit{The J. of Strategic Info. Sys.}, vol. 30, no. 3, pp. 101671, 2021.

  \item [{[P116]}] M. Liu, S. Hansen, and Q. Tu. ``Keeping the family together: Sustainability and modularity in community source development,'' \textit{Information and Organization}, vol. 30, no. 1, pp. 100274, 2020.

  \item [{[P117]}] K. M. Landgraf, R. Kakkar, M. Meigs, P. T. Jankauskas, T. T. H. Phan, V. N. Nguyen, D. T. Nguyen, T. T. Duong, T. H, Nguyen, and K. B. Bond. ``Open-source LIMS in Vietnam: The path toward sustainabilityand host country ownership,'' \textit{Intl. J. of Medical Informatics}, vol. 93, pp. 92-102, 2016.

  \item [{[P118]}] A. L. Zanatta, I. Steinmacher, L. S. Machado, C. R. De Souza, and R. Prikladnicki. ``Barriers faced by newcomers to software-crowdsourcing projects,'' \textit{IEEE Software}, vol. 34, no. 2, pp. 37-43, 2017.

  \item [{[P125]}] J. L. Cánovas Izquierdo, and J. Cabot. ``Enabling the Definition and Enforcement of Governance Rules in Open Source Systems,'' in \textit{Proc. of ICSE}. IEEE, 2015, pp. 505-514.

  \item [{[P126]}] M. A. Ferrario, W. Simm, S. Forshaw, A. Gradinar, M. T. Smith, and I. Smith. ``Values-First SE: Research Principles in Practice,'' in \textit{ICSE Companion Proc.}. IEEE, 2016, pp. 553-562.

  \item [{[P149]}] R. Govindaraju, Y. Y. Wibisono and A. Z. Sidiq. ``Critical processes in developing client-vendor relationship in the case of offshore IT/IS outsourcing,'' in \textit{2015 Intl. Conf. on Information Technology Systems and Innovation (ICITSI)}. IEEE, 2015, pp. 1-6.

  \item [{[P151]}] E. Jagroep, J. Broekman, J. M. E. Van Der Werf, P. Lago, S. Brinkkemper, L. Blom, and R. van Vliet. ``Awakening Awareness on Energy Consumption in Software Engineering,'' in \textit{ICSE Companion Proc.}. IEEE, 2017, pp. 76-85.

  \item [{[P156]}] L. Barroca, P. Gregory, K. Kuusinen, H. Sharp, and R. AlQaisi. ``Sustaining Agile Beyond Adoption,'' in \textit{Proc. of SEAA}. IEEE, 2018, pp. 22-25.

  \item [{[P165]}] R. Chitchyan, J. Noppen and I. Groher. ``What Can Software Engineering Do for Sustainability: Case of Software Product Lines,'' in \textit{2015 IEEE/ACM 5th Intl. Workshop on Product Line Approaches in Software Engineering}. IEEE, 2015, pp. 11-14.

  \item [{[P172]}] N. Rashid, S. U. Khan, H. U. Khan, and M. Ilyas. ``Green-Agile Maturity Model: An Evaluation Framework for Global Software Development Vendors,'' \textit{IEEE Access}, vol. 9, pp. 71868-71886. 2021.

  \item [{[P177]}] A. Sambhanthan, and V. Potdar. ``Waste management strategies for Software Development companies: An explorative text analysis of business sustainability reports,'' in \textit{2016 IEEE 14th Intl. Conf. on Software Engineering Research, Management and Applications (SERA)}. IEEE, 2016, pp. 179-184.

  \item [{[P180]}] I. Groher, and R. Weinreich. ``An interview study on sustainability concerns in software development projects,'' in \textit{Proc. of SEAA}. IEEE, 2017, pp. 350-358.

  \item [{[P183]}] B. Penzenstadler, S. Betz, C. C. Venters, R. Chitchyan, J. Porras, N. Seyff, L. Duboc, and C. Becker. ``Everything is INTERRELATED: Teaching software engineering for sustainability,'' in \textit{ICSE Companion Proc.)}. IEEE, 2018, pp. 153-162.

  \item [{[P186]}] I. Manotas, C. Bird, R. Zhang, D. Shepherd, C. Jaspan, C. Sadowski, L. Pollock, and J. Clause. ``An empirical study of practitioners' perspectives on green software engineering,'' in \textit{Proc. of ICSE}. IEEE, 2016, pp. 237-248.

  \item [{[P206]}] C. Ponsard, R. De Landtsheer, and F. Germeau. ``Building sustainable software for sustainable systems: case study of a shared pick-up and delivery service,'' in \textit{Proc. of the 6th Intl. Workshop on Green and Sustainable Software}. ACM, 2018, pp. 31-34.

  \item [{[P208]}] R. Chitchyan, C. Becker, S. Betz, L. Duboc. Penzenstadler, N. Seyff, and C. C. Venters ``Sustainability design in requirements engineering: state of practice,'' in \textit{ICSE Companion Proc.}. ACM, 2016, pp. 533-542.

  \item [{[P212]}] B. Penzenstadler, J. Mehrabi, and D. J. Richardson. ``Supporting physicians by RE4S: Evaluating requirements engineering for sustainability in the medical domain,'' in \textit{2015 IEEE/ACM 4th Intl. Workshop on Green and Sustainable Software}. IEEE, 2015, pp. 36-42.

  \item [{[P216]}] B. Penzenstadler, H. Femmer, and D. Richardson, D. ``Who is the advocate? Stakeholders for sustainability,'' in \textit{2013 2nd Intl. Workshop on Green and Sustainable Software (GREENS)}. IEEE, 2013, pp. 70-77.

  \item [{[P219]}] O. Cico, L. Jaccheri, and A. N. Duc. ``Software Sustainability in Customer-Driven Courses,'' in \textit{2021 IEEE/ACM Intl. Workshop on Body of Knowledge for Software Sustainability (BoKSS)}. IEEE, 2021, pp. 15-22.

  \item [{[P220]}] I. S. Brito, J. M. Conejero, A. Moreira, and J. Araújo. ``A concern-oriented sustainability approach,'' in \textit{Proc. of RCIS}. IEEE, 2018, pp. 1-12.
  
  \item [{[P243]}] J. Porras, C. C. Venters, B. Penzenstadler, L. Duboc, S. Betz, N. Seyff, S. Heshmatisafa and S. Oyedeji. ``How Could We Have Known? Anticipating Sustainability Effects of a Software Product,'' in \textit{Intl. Conf. on Software Business}. Springer, 2021, pp. 10-17.
\end{description}
\end{footnotesize}

\ifCLASSOPTIONcaptionsoff
  \newpage
\fi

\bibliographystyle{IEEEtran}

\balance

\bibliography{bib.bib}

\begin{thebibliography}{10}
\providecommand{\url}[1]{#1}
\csname url@samestyle\endcsname
\providecommand{\newblock}{\relax}
\providecommand{\bibinfo}[2]{#2}
\providecommand{\BIBentrySTDinterwordspacing}{\spaceskip=0pt\relax}
\providecommand{\BIBentryALTinterwordstretchfactor}{4}
\providecommand{\BIBentryALTinterwordspacing}{\spaceskip=\fontdimen2\font plus
\BIBentryALTinterwordstretchfactor\fontdimen3\font minus
  \fontdimen4\font\relax}
\providecommand{\BIBforeignlanguage}[2]{{%
\expandafter\ifx\csname l@#1\endcsname\relax
\typeout{** WARNING: IEEEtran.bst: No hyphenation pattern has been}%
\typeout{** loaded for the language `#1'. Using the pattern for}%
\typeout{** the default language instead.}%
\else
\language=\csname l@#1\endcsname
\fi
#2}}
\providecommand{\BIBdecl}{\relax}
\BIBdecl

\bibitem{brundtland1987report}
G.~H. Brundtland and {World Commission on Environment and Development},
  \emph{Report of the World Commission on Environment and Development: Our
  Common Future.}\hskip 1em plus 0.5em minus 0.4em\relax Oxford: Oxford
  University Press, 1987.

\bibitem{becker2023insolvent}
C.~Becker, \emph{Insolvent: How to Reorient Computing for Just
  Sustainability}.\hskip 1em plus 0.5em minus 0.4em\relax Boston, USA: MIT
  Press, 2023.

\bibitem{verdecchia2021green}
R.~Verdecchia, P.~Lago, C.~Ebert, and C.~de~Vries, ``Green it and green
  software,'' \emph{IEEE Software}, vol.~38, no.~6, pp. 7--15, 2021.

\bibitem{chitchyan2016sustainability}
R.~Chitchyan, C.~Becker, S.~Betz, L.~Duboc, B.~Penzenstadler, N.~Seyff, and
  C.~C. Venters, ``Sustainability design in requirements engineering: State of
  practice,'' in \emph{Proceedings of the 38th International Conference on
  Software Engineering Companion}.\hskip 1em plus 0.5em minus 0.4em\relax
  Association for Computing Machinery, 2016, p. 533–542.

\bibitem{becker2015sustainability}
C.~Becker, R.~Chitchyan, L.~Duboc, S.~Easterbrook, B.~Penzenstadler, N.~Seyff,
  and C.~C. Venters, ``Sustainability design and software: The karlskrona
  manifesto,'' in \emph{2015 IEEE/ACM 37th IEEE International Conference on
  Software Engineering}.\hskip 1em plus 0.5em minus 0.4em\relax IEEE, 2015, pp.
  467--476.

\bibitem{penzenstadler2013generic}
B.~Penzenstadler and H.~Femmer, ``A generic model for sustainability with
  process- and product-specific instances,'' in \emph{Proceedings of the 2013
  Workshop on Green in/by Software Engineering}.\hskip 1em plus 0.5em minus
  0.4em\relax Association for Computing Machinery, 2013, p. 3–8.

\bibitem{Markus2010Green}
M.~Dick, S.~Naumann, and N.~Kuhn, ``A model and selected instances of green and
  sustainable software,'' in \emph{What Kind of Information Society?
  Governance, Virtuality, Surveillance, Sustainability, Resilience},
  J.~Berleur, M.~D. Hercheui, and L.~M. Hilty, Eds.\hskip 1em plus 0.5em minus
  0.4em\relax Berlin, Heidelberg: Springer, 2010, pp. 248--259.

\bibitem{penzenstadler2012sustainability}
B.~Penzenstadler, V.~Bauer, C.~Calero, and X.~Franch, ``Sustainability in
  software engineering: A systematic literature review,'' in \emph{16th
  International Conference on Evaluation \& Assessment in Software Engineering
  (EASE 2012)}.\hskip 1em plus 0.5em minus 0.4em\relax IET, 2012, pp. 32--41.

\bibitem{penzenstadler2014systematic}
B.~Penzenstadler, A.~Raturi, D.~Richardson, C.~Calero, H.~Femmer, and
  X.~Franch, ``Systematic mapping study on software engineering for
  sustainability (se4s),'' in \emph{Proceedings of the 18th International
  Conference on Evaluation and Assessment in Software Engineering}, 2014, pp.
  1--14.

\bibitem{mourao2018green}
B.~C. Mour{\~a}o, L.~Karita, and I.~do~Carmo~Machado, ``Green and sustainable
  software engineering - a systematic mapping study,'' in \emph{Proceedings of
  the 17th Brazilian Symposium on Software Quality}, 2018, pp. 121--130.

\bibitem{marimuthu2017software}
C.~Marimuthu and K.~Chandrasekaran, ``Software engineering aspects of green and
  sustainable software: A systematic mapping study,'' in \emph{Proceedings of
  the 10th Innovations in Software Engineering Conference}, 2017, pp. 34--44.

\bibitem{garcia2018interactions}
G.~A. Garc{\'\i}a-Mireles, M.~{\'A}. Moraga, F.~Garc{\'\i}a, C.~Calero, and
  M.~Piattini, ``Interactions between environmental sustainability goals and
  software product quality: A mapping study,'' \emph{Information and Software
  Technology}, vol.~95, pp. 108--129, 2018.

\bibitem{gustavsson2020blinded}
J.~L. Gustavsson and B.~Penzenstadler, ``Blinded by simplicity: Locating the
  social dimension in software development process literature,'' in
  \emph{Proceedings of the 7th International Conference on ICT for
  Sustainability}, 2020, pp. 116--127.

\bibitem{wolfram2017sustainability}
N.~Wolfram, P.~Lago, and F.~Osborne, ``Sustainability in software
  engineering,'' in \emph{2017 Sustainable Internet and ICT for Sustainability
  (SustainIT)}.\hskip 1em plus 0.5em minus 0.4em\relax IEEE, 2017, pp. 1--7.

\bibitem{CALERO2017117}
C.~Calero and M.~Piattini, ``Puzzling out software sustainability,''
  \emph{Sustainable Computing: Informatics and Systems}, vol.~16, pp. 117--124,
  2017.

\bibitem{moises2018practices}
A.~C. Moises, A.~Malucelli, and S.~Reinehr, ``Practices of energy consumption
  for sustainable software engineering,'' in \emph{2018 Ninth International
  Green and Sustainable Computing Conference (IGSC)}.\hskip 1em plus 0.5em
  minus 0.4em\relax IEEE, 2018, pp. 1--6.

\bibitem{anwar2017}
H.~Anwar and D.~Pfahl, ``Towards greener software engineering using software
  analytics: A systematic mapping,'' in \emph{2017 43rd Euromicro Conference on
  Software Engineering and Advanced Applications (SEAA)}.\hskip 1em plus 0.5em
  minus 0.4em\relax IEEE, 2017, pp. 157--166.

\bibitem{calero2013systematic}
C.~Calero, M.~F. Bertoa, and M.~{\'A}. Moraga, ``A systematic literature review
  for software sustainability measures,'' in \emph{2013 2nd international
  workshop on green and sustainable software (GREENS)}.\hskip 1em plus 0.5em
  minus 0.4em\relax IEEE, 2013, pp. 46--53.

\bibitem{debbarma2016green}
T.~Debbarma and K.~Chandrasekaran, ``Green measurement metrics towards a
  sustainable software: A systematic literature review,'' in \emph{2016
  International Conference on Recent Advances and Innovations in Engineering
  (ICRAIE)}.\hskip 1em plus 0.5em minus 0.4em\relax IEEE, 2016, pp. 1--7.

\bibitem{salam2016developing}
M.~Salam and S.~U. Khan, ``Developing green and sustainable software: Success
  factors for vendors,'' in \emph{2016 7th IEEE International Conference on
  Software Engineering and Service Science (ICSESS)}.\hskip 1em plus 0.5em
  minus 0.4em\relax IEEE, 2016, pp. 1059--1062.

\bibitem{da2017has}
S.~da~Silva~Amorim, F.~S.~S. Neto, J.~D. McGregor, E.~S. de~Almeida, and C.~von
  Flach G.~Chavez, ``How has the health of software ecosystems been evaluated?
  a systematic review,'' in \emph{Proceedings of the 31st Brazilian symposium
  on software engineering}, 2017, pp. 14--23.

\bibitem{arksey2005scoping}
H.~Arksey and L.~O'Malley, ``Scoping studies: towards a methodological
  framework,'' \emph{International journal of social research methodology},
  vol.~8, no.~1, pp. 19--32, 2005.

\bibitem{ralph2022paving}
P.~Ralph and S.~Baltes, ``Paving the way for mature secondary research: The
  seven types of literature review,'' in \emph{Proceedings of The ACM Joint
  European Software Engineering Conference and Symposium on the Foundations of
  Software Engineering (ESEC/FSE 2022) Ideas, Visions and Reflections Track},
  2022.

\bibitem{baltes2022sampling}
S.~Baltes and P.~Ralph, ``Sampling in software engineering research: A critical
  review and guidelines,'' \emph{Empirical Software Engineering}, vol.~27,
  no.~4, pp. 1--31, 2022.

\bibitem{page2021prisma}
M.~J. Page, J.~E. McKenzie, P.~M. Bossuyt, I.~Boutron, T.~C. Hoffmann, C.~D.
  Mulrow, L.~Shamseer, J.~M. Tetzlaff, E.~A. Akl, S.~E. Brennan \emph{et~al.},
  ``The prisma 2020 statement: an updated guideline for reporting systematic
  reviews,'' \emph{Systematic Reviews}, vol.~10, no.~1, pp. 1--11, 2021.

\bibitem{stol2016grounded}
K.-J. Stol, P.~Ralph, and B.~Fitzgerald, ``Grounded theory in software
  engineering research: a critical review and guidelines,'' in
  \emph{Proceedings of the 38th International conference on software
  engineering}, 2016, pp. 120--131.

\bibitem{ralph2020standards}
P.~Ralph, ``Empirical standards for software engineering research,''
  \emph{arXiv preprint arXiv:2010.03525}, 2020.

\bibitem{finlayson2008qualitative}
K.~W. Finlayson and A.~Dixon, ``Qualitative meta-synthesis: a guide for the
  novice,'' \emph{Nurse Researcher}, vol.~15, no.~2, 2008.

\bibitem{nye2016origins}
E.~Nye, G.~Melendez-Torres, and C.~Bonell, ``Origins, methods and advances in
  qualitative meta-synthesis,'' \emph{Review of Education}, vol.~4, no.~1, pp.
  57--79, 2016.

\bibitem{fairclough2013critical}
N.~Fairclough, \emph{Critical Discourse Analysis: The Critical Study of
  Language}.\hskip 1em plus 0.5em minus 0.4em\relax New York, USA: Routledge,
  2013.

\bibitem{purvis2019three}
B.~Purvis, Y.~Mao, and D.~Robinson, ``Three pillars of sustainability: in
  search of conceptual origins,'' \emph{Sustainability science}, vol.~14,
  no.~3, pp. 681--695, 2019.

\bibitem{ungar2020resilience}
M.~Ungar and L.~Theron, ``Resilience and mental health: How multisystemic
  processes contribute to positive outcomes,'' \emph{The Lancet Psychiatry},
  vol.~7, no.~5, pp. 441--448, 2020.

\bibitem{holland1992complex}
J.~H. Holland, ``Complex adaptive systems,'' \emph{Daedalus}, vol. 121, no.~1,
  pp. 17--30, 1992.

\bibitem{hilty2015ict}
L.~M. Hilty and B.~Aebischer, ``Ict for sustainability: An emerging research
  field,'' in \emph{ICT Innovations for Sustainability}.\hskip 1em plus 0.5em
  minus 0.4em\relax Cham: Springer, 2015, pp. 3--36.

\bibitem{patterson2021carbon}
D.~Patterson, J.~Gonzalez, Q.~Le, C.~Liang, L.-M. Munguia, D.~Rothchild, D.~So,
  M.~Texier, and J.~Dean, ``Carbon emissions and large neural network
  training,'' \emph{arXiv preprint arXiv:2104.10350}, 2021.

\bibitem{edholm2017crunch}
H.~Edholm, M.~Lidstr{\"o}m, J.-P. Stegh{\"o}fer, and H.~Burden, ``Crunch time:
  The reasons and effects of unpaid overtime in the games industry,'' in
  \emph{2017 IEEE/ACM 39th International Conference on Software Engineering:
  Software Engineering in Practice Track (ICSE-SEIP)}.\hskip 1em plus 0.5em
  minus 0.4em\relax IEEE, 2017, pp. 43--52.

\bibitem{nelson2007project}
R.~R. Nelson, ``It project management: Infamous failures, classic mistakes, and
  best practices.'' \emph{MIS Quarterly Executive}, vol.~6, no.~2, pp. 67--78,
  2007.

\bibitem{sonnentag1994stressor}
S.~Sonnentag, F.~C. Brodbeck, T.~Heinbokel, and W.~Stolte, ``Stressor-burnout
  relationship in software development teams,'' \emph{Journal of Occupational
  and Organizational Psychology}, vol.~67, no.~4, pp. 327--341, 1994.

\end{thebibliography}

\end{document}